# Density Functional Theory and Deep-learning to Accelerate Data Analytics in Scanning Tunneling Microscopy

Kamal Choudhary[1], Kevin F. Garrity[1], Charles Camp[1], Sergei V. Kalinin[2], Rama Vasudevan[2], Maxim Ziatdinov[2], Francesca Tavazza[1]

1. Material Measurement Laboratory, National Institute of Standards and Technology, Gaithersburg, MD 20899, USA.

2. Center for Nanophase Materials Sciences, Oak Ridge National Laboratory, Oak Ridge TN 37831, USA.

## Abstract

We introduce the first systematic database of scanning tunneling microscope (STM) images obtained using density functional theory (DFT) for two-dimensional (2D) materials, calculated using the Tersoff-Hamann method. It currently contains data for 716 exfoliable 2D materials. Examples of the five possible Bravais lattice types for 2D materials and their Fourier-transforms are discussed. All the computational STM images generated in this work will be made available on the JARVIS-DFT website (https://www.ctcms.nist.gov/~knc6/JVASP.html). We find excellent qualitative agreement between the computational and experimental STM images for selected materials. As a first example application of this database, we train a convolution neural network (CNN) model to identify Bravais lattices from the STM images. We believe the model can aid high-throughput experimental data analysis. These computational STM images can directly aid the identification of phases, analyzing defects and lattice-distortions in experimental STM images, as well as be incorporated in the autonomous experiment workflows.

**Corresponding author**: kamal.choudhary@nist.gov

**Introduction**

Since the invention of the scanning tunneling microscope (STM)[1], this technique has become an essential tool for characterizing material surfaces and adsorbates. In addition to providing atomic insights, STM has been proven useful for characterizing electronic structure, shapes of molecular orbitals, and vibrational and magnetic excitations[2,3]. It can also be used for manipulating adsorbates and adatoms, and for catalysis and quantum information processing applications[3-8]. Quantum mechanics-based density functional theory (DFT) has often been used to produce virtual STM images for these applications [9,10]. However, a systematic database of such computational STM data is still lacking. As DFT-STM images are constructed using defect-free materials, they provide standard reference images (SRI) that are useful to aid in identifying phases[11,12], analyzing defects[13,14] and quantifying lattice-distortions[15] in experimental STM images. A DFT-STM database is therefore essential to provide a direct link between atomic positions and images, which can aid experimental analysis. Moreover, the orbital-projected density of states available in our database can help explain which atoms and orbitals contribute to an experimental STM image. Finally, a computational database can provide an accurate training set for developing machine learning (ML) models to rapidly analyze experimental STM images.

STM imaging is particularly well-suited to studying two-dimensional (2D) materials, such as graphene[16], $MoS_2$[17], $NbSe_2$[18], $WSe_2$[19], $WTe_2$[20], FeSe[21], black-phosphrous[22,23] and SnSe[24]. 2D materials[25,26] are a special class of materials with diverse areas of application, such as sub-micron level electronics[27], flexible and tunable electronics[28], superconductivity[29], photo-voltaics[30], water-purification[31], sensors[32], thermal-management[33], energy-storage[34], medicine[35], quantum dots[36,37] and composites[38-40]. The surfaces of 2D materials are unique because they lack dangling bonds, allowing them to be exfoliated. This property makes them ideal candidates for building a database of computational STMs images because they don't require thick slabs perpendicular to the surface, which are computationally expensive to simulate accurately, and they do not have surface reconstructions. The generation of STM images for perfect systems is an initial step, and we will extend this project to include defective systems in the future.

In this work, we use DFT to generate STM images of exfoliable 2D materials. We use the recently developed JARVIS-DFT database (https://www.ctcms.nist.gov/~knc6/JVASP.html) and select 2D materials with exfoliation energy less than 200 meV/atom. The JARVIS-DFT database contains about 40000 bulk and 1000 two-dimensional materials with their DFT-computed structural, energetic[41], elastic[42], optoelectronic[43], thermoelectric[44], piezoelectric, dielectric, infrared[45], solar-efficiency[46], and topological[47] properties. We note that there are several factors that can influence the appearance of experimental or DFT-based STM image predictions, such as the STM-tip material, bias voltage, and the scanning mode, *i.e.* constant-height mode (CHM) vs. constant current mode (CCM). Similarly, there are several methods for simulating STM images using DFT, including Bardeen[48], Tersoff-Hamman[49] and Chen[3] methods. Here, we present results for constant height and constant current DFT-STM images computed using the Tersoff-Hamann approach[49], which assumes a non-functionalized (*s*-wave) STM tip. The ML model training is based on CHM images. The DFT-STM database currently contains images for 716 materials, with additional computations ongoing. All the DFT-STM data will be uploaded into the JARVIS-DFT database.

To leverage artificial intelligence methods[50] to automatically characterize STM images, we use the computational STM images to train a convolution neural network ML classification model for Bravais-lattices. This model is able to quickly classify STM images into the five lattice classes (square, hexagon, rhombus/centered-rectangle, rectangle and parallelogram/oblique) that are possible for 2D systems. Such classifications are of importance, for example when dealing with phase transitions[51]. Ideally one would use an information-theoretic approach, as opposed to deep learning, to enable space group determination with uncertainty quantification, as demonstrated by Moeck[52]. However, a pre-screening step can be rapidly accomplished with a suitably trained neural network as shown here. Later, these computational STM trained models can be integrated with experiments for active learning processes.[50]

## Results and discussion

We simulate computational STM images of 716 exfoliable materials ($E_f < 200$ meV/atom) using the Tersoff-Hamann approach. We compare computational STM images with those from experiments for graphene[16], 2H-MoS$_2$[17], 2H-NbSe$_2$[18], 2H-WSe$_2$[19], 1T'-WTe$_2$[20], FeSe[21], black-P[22,23], SnSe[24], Bismuth[58,59]. Qualitatively, we observe that the patterns in the computational and experimental STMs are very similar (see the supplementary information, Fig. S1). Note that we are able to predict the STM for 2D very well because they lack dangling bonds. Such images with non-vdW systems such as Si(111)[60] would require bigger simulation cells to accommodate reconstructions and many layers to converge the calculation and allow comparison with experiments.

The DFT-STM can be used for distinguishing phases such as the 2D-monolayer MoTe$_2$ 2H (JVASP-670) and 1T' (JVASP-673) phases, as shown in constant height positive bias conditions in Fig. 1. The 2H-phase is semiconducting material with hexagonal symmetry, as is evident from the crystal structure in Fig. 1a. The positive +0.5 eV bias constant height image of this structure is shown in Fig. 1b. The electronic states in this range are dominated by Mo (*d*-orbital) states, hence the brighter spots in the STM are dominated by Mo *d*-orbitals, which can be understood by analyzing the projected density of states (Fig. S2). As shown in Fig. 1c, the fast Fourier transform (FT) of the simulated STM image in Fig. 1b shows hexagonal symmetry. Similarly, the crystal structure, STM image and FT of rectangular 1T'-MoTe$_2$ is shown in Fig. 1d, Fig. 1e and Fig. 1f respectively. We note that the FT of the STM image of a rectangular system with multi-atom cell is not a simple rectangle. We show examples of variation of height in Å and current in arbitrary units in Fig. 1g,h and i for 2H-MoTe$_2$. The constant height for 2H-MoTe$_2$ in Fig. 1b is for 3.7 Å while that in Fig. 1g is for 7.4 Å with respect to the highest atom in the cell. Clearly, the hexagonal patterns remain the same, but the structure around the atoms changes due to the change in height. This is because as we move the hypothetical STM tip, we probe different layers of charge density. Similarly, we show the current variation based STM images for 0.01 and 0.05 a.u.$^{-3}$ eV$^{-1}$ in Fig. h and i. Note that it is difficult to quantitatively compare the computational and experimental STM images, because the tunneling-current is critically dependent on the specific experimental setup.

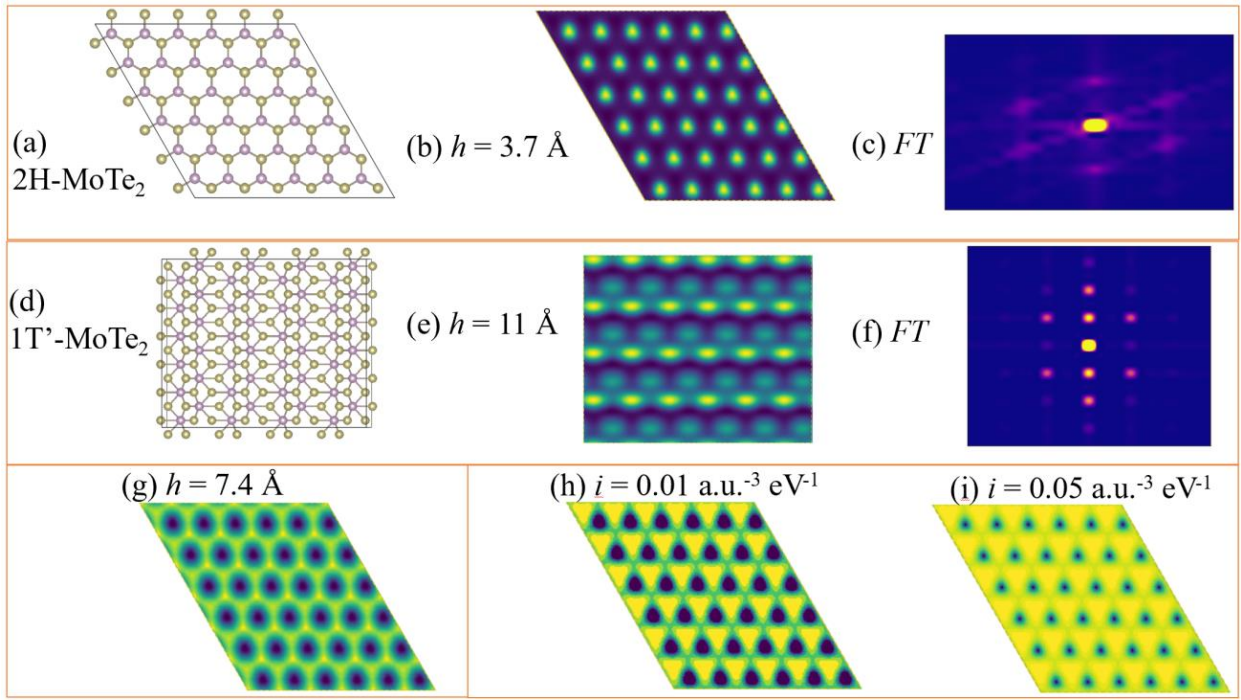

*Fig. 1 STM images of MoTe$_2$ in 2H and 1T' phases, their fast Fourier transform (FT), height and current dependence. a) crystal structure of 2H-MoTe$_2$ (JVASP-667) b) its constant-height STM image at the height of 3.7 Å, c) FT of the STM image, d) crystal structure of 1T'-MoTe$_2$ (JVASP-673) e) its STM image at the height of 11 Å, f) FT of the STM image, g) constant-height STM image of 2H-MoTe$_2$ at the height of 7.4 Å to compare with Fig. b, h) constant current image for 2H-MoTe$_2$ at constant current* 0.01 a.u.$^{-3}$ eV$^{-1}$, *i) constant current image for 2H-MoTe$_2$ at constant current* 0.05 a.u.$^{-3}$ eV$^{-1}$.

Based on lattice parameter information in 2D plane, the 2D materials lattices can be classified in 5 types: hexagon, square, rectangle, rhombus/centered-rectangle, and parallelograms/oblique. We classify all the 2D materials in our database, with the distribution shown in Fig. 2a. Most of the 2D materials in our database are hexagonal, followed by rectangular and square lattices. In Fig. 2, we give examples of materials in each lattice type, in each case showing the atomic positions, a constant height STM image, and the fast Fourier transform (FT) of the STM image. An example of hexagonal lattice is shown in Fig. 2b graphene (JVASP-667). It is one of the most widely investigated 2D materials. The STM positive bias image for graphene is shown in Fig. 2c. An FT of the image 2c is shown in Fig. 2d. It is clear from Fig. 2d that that there is hexagonal pattern due to hexagonal symmetry in graphene. Similarly, for the square lattice example, FeTe (JVASP-6667), the crystal structure, STM, and FT are shown in Fig. 2e-g. Fe *d*-states mainly contributes to the STM image in Fig. 2f. The FT of this image shows square-like patterns in Fig. 2g. Similarly, Fig. 2h gives the crystal structure of VClO (JVASP-8933), and its STM and FT show a rectangular pattern (Fig. 2i and Fig. 2j). AuI (JVASP-6187) has a rhombus structure, as shown in Fig. 2k. The lattice constants are 4.274 Å and the angle between them is 93.2 degrees. The Au *d*-orbitals contribute most to the STM image. The atomic and orbital projected density of systems for all the systems here is given in the supplementary information (Fig. S3) and the respective webpages for each material. The FT in the Fig. 2 shows noticeable blur, which can be caused by the truncation of the infinite slab to a finite image. Fig. 2n shows As$_2$Se$_3$ (JVASP-13544), an example material with a parallelogram unit cell with lattice constants of 4.4 and 12.9 Å and an angle of 109.9 degrees. The FT of the STM is difficult to interpret.

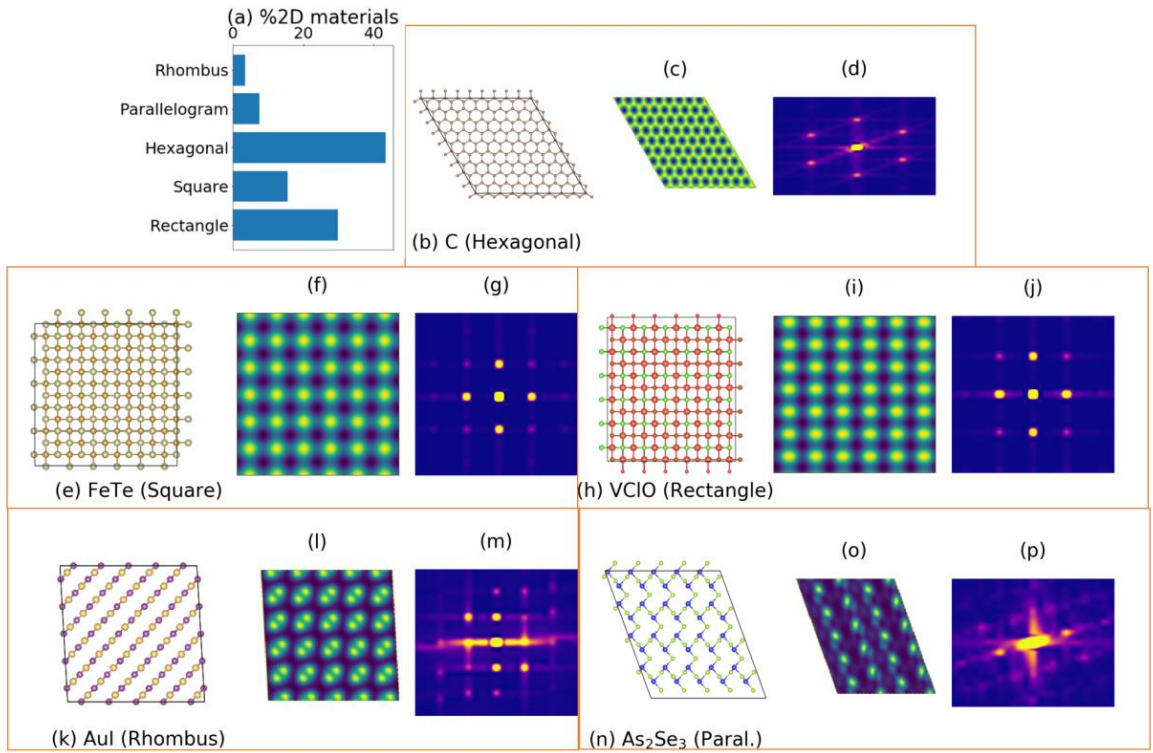

*Fig. 2 a) 2D lattice type distribution in the database. b-d) crystal structure, constant height STM (CHS) and FT for graphene. Similar images for e-g) FeTe, h-j) VClO, k-m) AuI, n-p) $As_2Se_3$.*

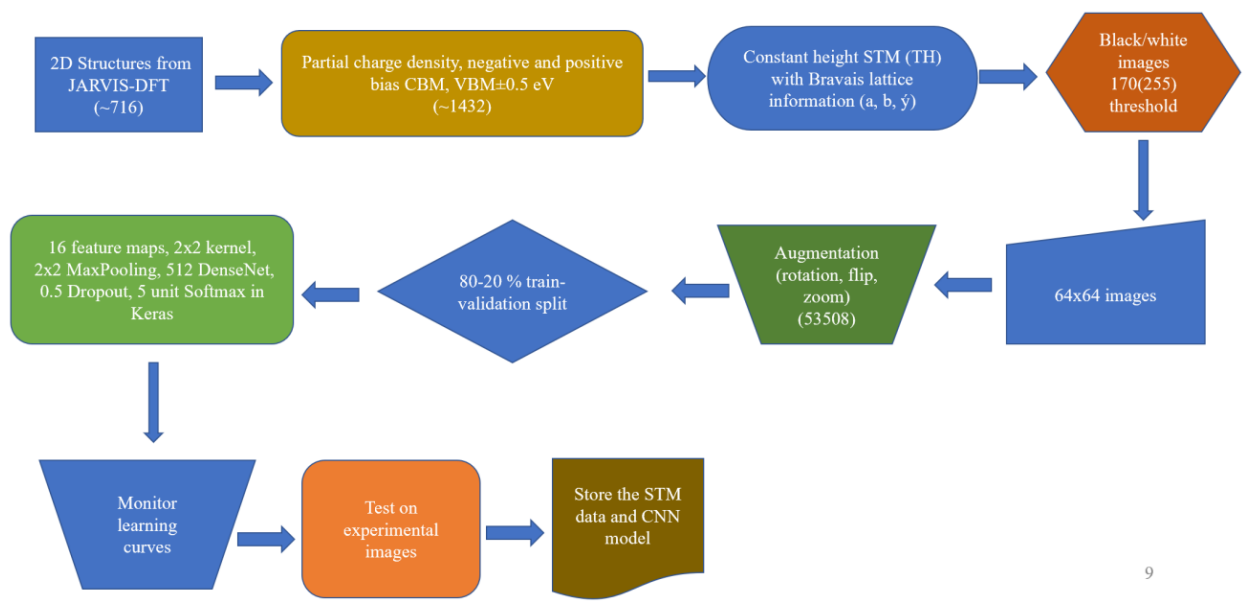

*Fig. 3 Flow-chart showing the steps involved in the machine learning process.*

Having prepared our database, we now train a ML model following the flow-chart in Fig. 3. First, for all the original images, we generate augmented images with rotation, slide, zoom-in and zoom-out operations randomly applied such that each class has at least 10000 images. This leads to 53508 images. We use multi-layer network with one convolution layer (16 feature maps, with 2x2 kernel) activated by a rectified linear unit (ReLU), one max-pooling layer, one fully-connected 512-nodes layer with ReLU activations, and a fully-connected softmax layer with five outputs. Since the entire dataset is too big to feed to the GPU memory at once, we divide it in multiple smaller batches. The total number of training examples present in a single batch (batch size) is 32 for our NN model. We have 20% dropout before the softmax layer to avoiding overfitting. We use ADAM stochastic optimization method for gradient descent with 'sparse categorical crossentropy' as loss function. We have 80-20% split for train and validation data.

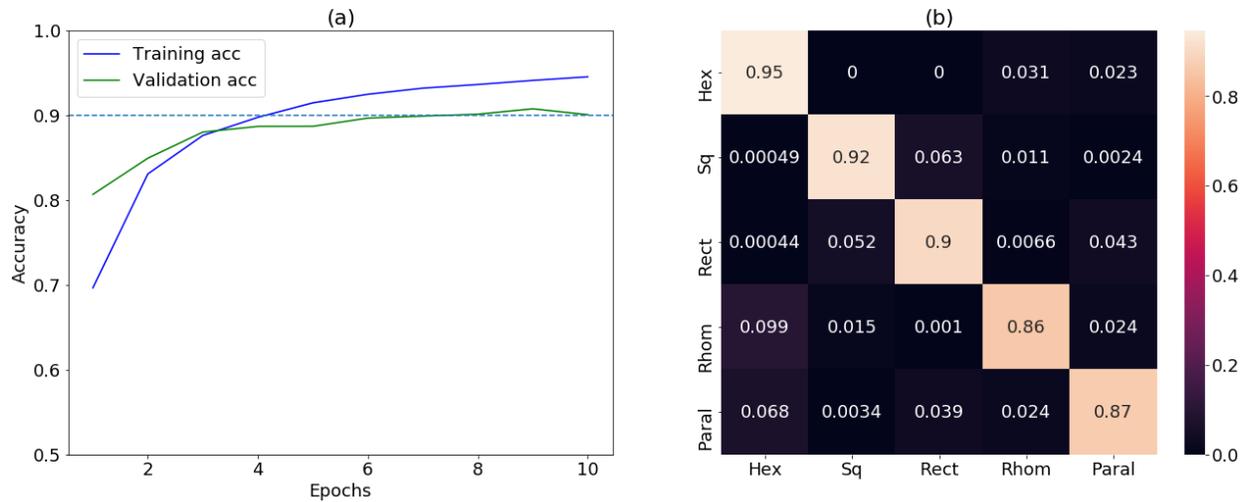

*Fig. 4 Performance of the machine learning model. a) learning curve, b) confusion matrix.*

*Table. 1 Classification report of classifying 2D constant-height STM images into lattice-types.*

| Lattice | Precision | Recall | F1-score |
|---|---|---|---|
| **Hexagonal** | 0.87 | 0.95 | 0.91 |
| **Square** | 0.92 | 0.92 | 0.92 |
| **Rectangle** | 0.91 | 0.90 | 0.90 |
| **Rhombus** | 0.91 | 0.86 | 0.89 |
| **Parallelogram** | 0.90 | 0.87 | 0.88 |
| **Accuracy** | | | 0.90 |

In Fig. 4 we show the convolution neural network training and the learning curves for the deep learning model. We monitor the learning curve as in Fig. 4a. We see that after the 4$^{th}$ epoch the training and validation accuracy curves begin to diverge, so we stop further training. We obtain 90% accuracy on the validation set. The difference between the training and the validation curve is small, implying low overfitting. Applying the training model on the validation data gives the confusion matrix shown in Fig. 4b. We also provide precision, recall and F1 scores in Table 1. Clearly, all the scores are more than 0.5, indicating that the model performs much better than a random guessing model. Note that although the accuracy is a measure of the overall model, it is important to investigate the prediction accuracy for each class of the model. A confusion matrix with high diagonal element values signifies high accuracy. It is clear from the Fig. 4b that the model performs excellently for hexagonal and square lattices, and less well for the rectangle, rhombus and parallelogram lattices. Moving beyond simulated STM images, as an initial validation, we apply the model to nine experimental images. We find that the model predicts the correct class for seven of them (see supplementary information, Fig. S1). Performing a more systemic analysis of our model's accuracy on experimental images would require a database of hundreds of experimental images, and such a database is currently not available. We hope this work will spur the development of such a database.

## Conclusions

We introduce the first systematic database of scanning tunneling microscope (STM) images obtained using density functional theory (DFT) for two-dimensional (2D) materials. Specifically, the database is constructed using the Tersoff-Hamann method for constant-height images. It currently contains data for 716 exfoliable 2D materials from the JARVIS-DFT database. These STM images can directly aid the experimental analysis of STM results, as they provide ideal reference images. Examples of the five possible Bravais lattice types for 2D materials are discussed, and the images for all the materials are available on the website. We validated the computational STM technique by comparing our results to several experimental images. Finally, in order to help automate the tedious analysis of experimental STM images, we used our DFT-STM image database to train a high-accuracy convolution neural network (CNN) model to accelerate image characterization.

## Methods

All DFT calculations are carried out with Vienna *ab initio* simulation package (VASP)[53,54] using projected augmented wave (PAW) formalism and using vdW-DF-OptB88 functional[55]. All the machine learning training are carried using Keras with TensorFlow backend[56]. Note that commercial software is identified to specify procedures, and such identification does not imply recommendation by the National Institute of Standards and Technology. The k-point and plane-wave cut-off convergence for each material are obtained using the workflow detailed in Ref.[57]. The 2D materials are provided with at least 20 Å vacuum in the *z*-direction to avoid self-interactions. The force and energy convergence for DFT self-consistent calculations are $10^{-6}$ eV and 0.001 eV/Å respectively. The surface charge and probability densities are calculated by integrating the local density of states function (ILDOS) over an energy range of ±0.5 eV from the conduction band minima (CBM) to Fermi energy ($E_F$) and valence band maxima (VBM) to Fermi energy ($E_F$). The STM images are calculated using Tersoff-Hamann approach[49]:

$$I_\mu \propto \rho(r, E_F) \equiv \sum_\mu |\psi_\mu(r)|^2 \delta(\varepsilon_\mu - E_F) \qquad (1)$$

where $I_\mu$ is the tunneling current, $\psi_\mu$ the eigenvector and $\varepsilon_\mu$ the eigenenergy due to the state μ and $E_F$ is the Fermi-energy. All the STM images are made at least 20 Å long in the *xy* plane by repeating the primitive unit cell. For constant-current images, we identify iso-surfaces that have a constant ILDOS. The height of these iso-surfaces at each *xy*-coordinates produces the images.

We simplify the constant-height STM images using a black/white color-scheme and choose a pixel value of 170 (out of maximum 255) for finding atomic features. Based on the lattice-parameters and angles the 2D materials can be classified in five classes: 1) hexagonal, 2) square, 3) rhombus/centered-rectangle, 4) rectangle, 5) parallelograms/oblique. Deep-learning image recognition tasks typically require thousands of training images. To increase the size of our training set, we use several commonly applied image augmentations: random rotations, flipping, zooming in and zooming out. We apply augmentations until all the five classes have at least 10000 images. We train-test the images using an 80-20% split in such a way that both the train and test types have an equal proportion of all the five classes of lattice types. During the training, we monitor the train-validation curve (discussed later) to avoid overfitting. We use accuracy, precision, recall, and F1-score to measure the overall and individual class performances. The precision is the ratio $t_p / (t_p + f_p)$ where $t_p$ is the number of true positives and $f_p$ the number of false positives. The recall is the ratio $t_p / (t_p + f_n)$ where $t_p$ is the number of true positives and $f_n$ the number of false negatives. The recall is intuitively the ability of the classifier to find all the positive samples. The F-1 score can be interpreted as a weighted harmonic mean of the precision and recall, where an F-1 score reaches its best value at 1 and worst score at 0. Both the model and the associated dataset will be made publicly available soon at the JARVIS-DFT website.

## Data availability

The electronic structure data will be made available at the JARVIS-DFT website: https://www.ctcms.nist.gov/~knc6/JVASP.html and http://jarvis.nist.gov .

## Contributions

KC developed the workflow, carried out the DFT calculations and trained the machine learning model. KC and KG analyzed the DFT data. CC, SK, RV, MZ helped in the machine learning training. RV, SK and MZ helped in the experimental validation. All contributed in writing the manuscript.

**Competing interests**

The authors declare no competing interests.

**References**

1   Binnig, G., Rohrer, H., Gerber, C. & Weibel, E. Surface studies by scanning tunneling microscopy. Phys. Rev. Lett. **49**, 57 (1982).
2   Mugarza, A. *et al.* Spin coupling and relaxation inside molecule–metal contacts. Nat. Comm. **2**, 490 (2011).
3   Chen, C. J. *Introduction to scanning tunneling microscopy*. Vol. 4 (Oxford University Press on Demand, 1993).
4   Gross, L. *et al.* High-resolution molecular orbital imaging using a p-wave STM tip. Phys. Rev. Lett. **107**, 086101 (2011).
5   Eigler, D. M. & Schweizer, E. K. Positioning single atoms with a scanning tunnelling microscope. Nature **344**, 524 (1990).
6   Stipe, B., Rezaei, M. & Ho, W. Single-molecule vibrational spectroscopy and microscopy. Science **280**, 1732-1735 (1998).
7   Hirjibehedin, C. F. *et al.* Large magnetic anisotropy of a single atomic spin embedded in a surface molecular network. Science **317**, 1199-1203 (2007).
8   Yang, K. *et al.* Coherent spin manipulation of individual atoms on a surface. Science **366**, 509-512 (2019).
9   Barth, J., Brune, H., Ertl, G. & Behm, R. Scanning tunneling microscopy observations on the reconstructed Au (111) surface: Atomic structure, long-range superstructure, rotational domains, and surface defects. Phys. Rev. B **42**, 9307 (1990).
10  Magonov, S. N. & Whangbo, M.-H. *Surface analysis with STM and AFM: experimental and theoretical aspects of image analysis*. (John Wiley & Sons, 2008).
11  Poirier, G. *et al.* Identification of the facet planes of phase I TiO2 (001) rutile by scanning tunneling microscopy and low energy electron diffraction. J. Vac. Sci. Tech. B **10**, 6-15 (1992).
12  M. Ziatdinov, A. Maksov, S. V. Kalinin, Learning Surface Molecular Structures via Machine Vision, npj Computational Materials 3, 31 (2017).
13  Vancsó, P. *et al.* The intrinsic defect structure of exfoliated MoS2 single layers revealed by Scanning Tunneling Microscopy. Sci. Rep. **6**, 29726 (2016).

14  Liu, H. *et al.* Line and point defects in MoSe2 bilayer studied by scanning tunneling microscopy and spectroscopy. ACS Nano **9**, 6619-6625 (2015).
15  Dubout, Q. *et al.* Giant apparent lattice distortions in STM images of corrugated sp2-hybridised monolayers. New J. Phys. **18**, 103027 (2016).
16  Li, G., Luican, A. & Andrei, E. Y. Scanning tunneling spectroscopy of graphene on graphite. Phys. Rev. Lett. **102**, 176804 (2009).
17  Mills, A. *et al.* Ripples near edge terminals in MoS2 few layers and pyramid nanostructures. App. Phys. Lett. **108**, 081601 (2016).
18  Wang, J. *et al.* A variable-temperature scanning tunneling microscope operated in a continuous flow cryostat. Rev. Sci. Instr. **90**, 093702 (2019).
19  Liu, H. *et al.* Molecular-beam epitaxy of monolayer and bilayer WSe2: a scanning tunneling microscopy/spectroscopy study and deduction of exciton binding energy. 2D Maters. **2**, 034004 (2015).
20  Jia, Z.-Y. *et al.* Direct visualization of a two-dimensional topological insulator in the single-layer 1 T'− WT e 2. Phys. Rev. B **96**, 041108 (2017).
21  Song, C.-L. *et al.* Direct observation of nodes and twofold symmetry in FeSe superconductor. Science **332**, 1410-1413 (2011).
22  Kumar, A. *et al.* STM study of exfoliated few layer black phosphorus annealed in ultrahigh vacuum. 2D Maters. **6**, 015005 (2018).
23  Kiraly, B., Hauptmann, N., Rudenko, A. N., Katsnelson, M. I. & Khajetoorians, A. A. Probing single vacancies in black phosphorus at the atomic level. Nano Lett. **17**, 3607-3612 (2017).
24  Duvjir, G. *et al.* Origin of p-type characteristics in a SnSe single crystal. App. Phys. Lett. **110**, 262106 (2017).
25  Xu, M., Liang, T., Shi, M. & Chen, H. Graphene-like two-dimensional materials. Chem Rev. **113**, 3766-3798 (2013).
26  Choudhary, K., Kalish, I., Beams, R. & Tavazza, F. High-throughput Identification and Characterization of Two-dimensional Materials using Density functional theory. Scientific Reports **7**, 5179 (2017).
27  Fiori, G. *et al.* Electronics based on two-dimensional materials. *Nature nanotechnology* **9**, 768-779 (2014).
28  Akinwande, D., Petrone, N. & Hone, J. Two-dimensional flexible nanoelectronics. *Nat Comm.* **5** (2014).
29  Navarro-Moratalla, E. & Jarillo-Herrero, P. Two-dimensional superconductivity: The Ising on the monolayer. *Nature Physics* **12**, 112-113 (2016).
30  Bubnova, O. 2D materials: Hybrid interfaces. *Nat Nano*, 16, 497 (2016).
31  Dervin, S., Dionysiou, D. D. & Pillai, S. C. 2D nanostructures for water purification: graphene and beyond. *Nanoscale* (2016).
32  Cui, S. *et al.* Ultrahigh sensitivity and layer-dependent sensing performance of phosphorene-based gas sensors. *Nature communications* **6** 8632 (2015).
33  Lee, M.-J. *et al.* Thermoelectric materials by using two-dimensional materials with negative correlation between electrical and thermal conductivity. *Nature Communications* **7** 12011 (2016).
34  Zhang, X., Hou, L., Ciesielski, A. & Samorì, P. 2D Materials Beyond Graphene for High-Performance Energy Storage Applications. *Advanced Energy Materials* (2016).
35  Boland, C. S. *et al.* Sensitive, high-strain, high-rate bodily motion sensors based on graphene–rubber composites. *ACS nano* **8**, 8819-8830 (2014).
36  Wang, X., Sun, G., Li, N. & Chen, P. Quantum dots derived from two-dimensional materials and their applications for catalysis and energy. *Chemical Society Reviews* **45**, 2239-2262 (2016).

37	Chakraborty, C., Kinnischtzke, L., Goodfellow, K. M., Beams, R. & Vamivakas, A. N. Voltage-controlled quantum light from an atomically thin semiconductor. *Nature nanotechnology* **10**, 507-511 (2015).
38	Castellanos-Gomez, A. Why all the fuss about 2D semiconductors? *Nat Photon* **10**, 202-204, doi:10.1038/nphoton.2016.53 (2016).
39	Flat talk. *Nat Photon* **10**, 205-206, (2016).
40	Rodenas, T. *et al.* Metal–organic framework nanosheets in polymer composite materials for gas separation. *Nature materials* **14**, 48-55 (2015).
41	Choudhary, K., Kalish, I., Beams, R. & Tavazza, F. High-throughput Identification and Characterization of Two-dimensional Materials using Density functional theory. *Scientific Reports* **7**, 5179 (2017).
42	Choudhary, K., Cheon, G., Reed, E. & Tavazza, F. Elastic properties of bulk and low-dimensional materials using van der Waals density functional. Physical Review B **98**, 014107 (2018).
43	Choudhary, K. *et al.* Computational screening of high-performance optoelectronic materials using OptB88vdW and TB-mBJ formalisms. *Scientific data* **5**, 180082 (2018).
44	Choudhary, K., Garrity, K. & Tavazza, F. Data-driven Discovery of 3D and 2D Thermoelectric Materials.  arXiv:1906.06024 (2019).
45	Choudhary, K. *et al.* High-throughput Density Functional Perturbation Theory and Machine Learning Predictions of Infrared, Piezoelectric and Dielectric Responses.  arXiv:1910.01183 (2019).
46	Choudhary, K. *et al.* Accelerated Discovery of Efficient Solar Cell Materials Using Quantum and Machine-Learning Methods. *Chemistry of Materials*, 31, 15, 5900 (2019).
47	Choudhary, K., DeCost, B. & Tavazza, F. Machine learning with force-field inspired descriptors for materials: fast screening and mapping energy landscape. Physical Review Materials. 2 083801 (2018).
48	Bardeen, J. Tunnelling from a many-particle point of view. Phys. Rev. Lett. **6**, 57 (1961).
49	Tersoff, J. & Hamann, D. R. Theory of the scanning tunneling microscope. Phys. Rev. B **31**, 805 (1985).
50	Vasudevan, R. K. *et al.* Materials science in the artificial intelligence age: high-throughput library generation, machine learning, and a pathway from correlations to the underpinning physics. MRS Comm. 9, 3, 821 (2019).
51	Vasudevan, R. K. *et al.* Mapping mesoscopic phase evolution during E-beam induced transformations via deep learning of atomically resolved images. npj Comp. Maters. **4**, 30 (2018).
52	Moeck, P. Towards generalized noise-level dependent crystallographic symmetry classifications of more or less periodic crystal patterns. Symm. **10**, 133 (2018).
53	Kresse, G. & Furthmüller, J. Efficient iterative schemes for ab initio total-energy calculations using a plane-wave basis set. Phys. Rev. B **54**, 11169 (1996).
54	Kresse, G. & Furthmüller, J. Efficiency of ab-initio total energy calculations for metals and semiconductors using a plane-wave basis set. Comp. Mat. Sci.  **6**, 15-50 (1996).
55	Klimeš, J., Bowler, D. R. & Michaelides, A. J. Chemical accuracy for the van der Waals density functional. J. Phys. Cond. Matt. **22**, 022201 (2009).
56	Abadi, M. *et al.* in *12th {USENIX} Symposium on Operating Systems Design and Implementation ({OSDI} 16).*  265-283.
57	Choudhary, K. & Tavazza, F. Convergence and machine learning predictions of Monkhorst-Pack k-points and plane-wave cut-off in high-throughput DFT calculations. Computational Materials Science  **161**, 300-308 (2019).
58	Sk, R., Deshpande, A. & Engineering. Unveiling the emergence of functional materials with STM: metal phthalocyanine on surface architectures. Molecular Systems Design & Engineering  (2019).

59  Song, F. *et al.* Low-temperature growth of bismuth thin films with (111) facet on highly oriented pyrolytic graphite. ACS Appl. Mater. Interfaces **7**, 8525-8532 (2015).
60  Smeu, M., Guo, H., Ji, W. & Wolkow, R. A.. Electronic properties of Si (111)-7× 7 and related reconstructions: Density functional theory calculations. Phys. Rev. B **85**, 195315 (2012).

**Supplementary info: Density Functional Theory and Deep-learning to Accelerate Data Analytics in Scanning Tunneling Microscopy**

Kamal Choudhary[1], Kevin F. Garrity[1], Charles Camp[1], Sergei V. Kalinin[2], Rama Vasudevan[2], Maxim Ziatdinov[2], Francesca Tavazza[1]

1. Material Measurement Laboratory, National Institute of Standards and Technology, Gaithersburg, MD 20899, USA.

2. Center for Nanophase Materials Sciences, Oak Ridge National Laboratory, Oak Ridge TN 37831, USA.

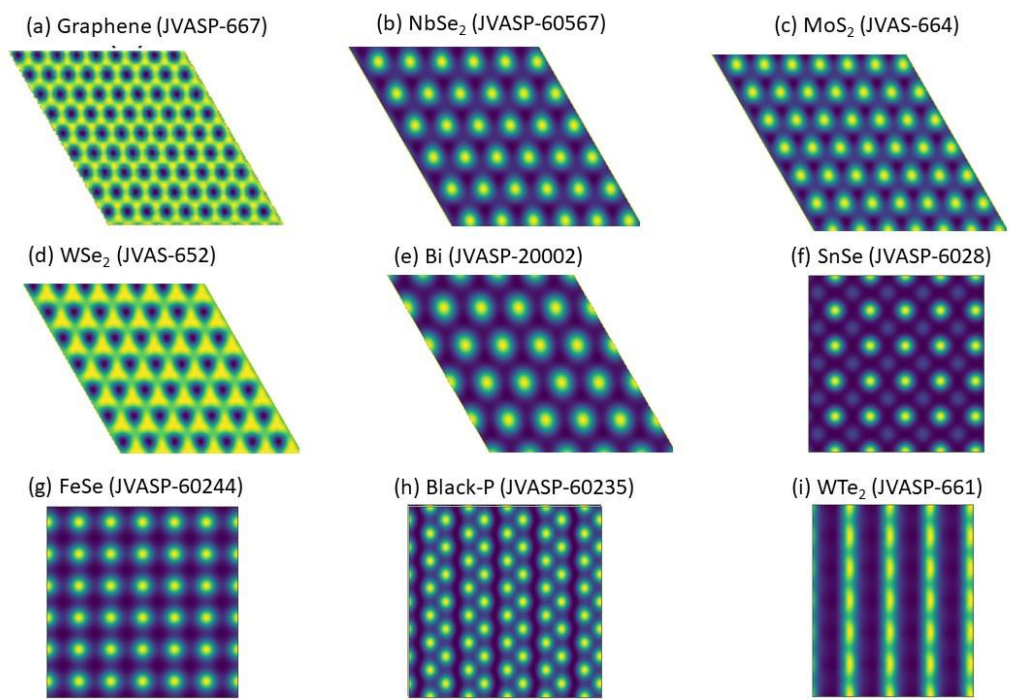

*Fig. S1 STM images for which we qualitatively compared computational STM images.* We compare computational STM images with that of experiments for graphene[16], 2H-MoS$_2$[17], 2H-NbSe$_2$[18], 2H-WSe$_2$[19], 1T'-WTe$_2$[20], FeSe[21], black-P[22,23], SnSe[24], Bismuth surface[58,59]. *The JVASP identifiers can be used to visualize the detailed webpage for each material.*

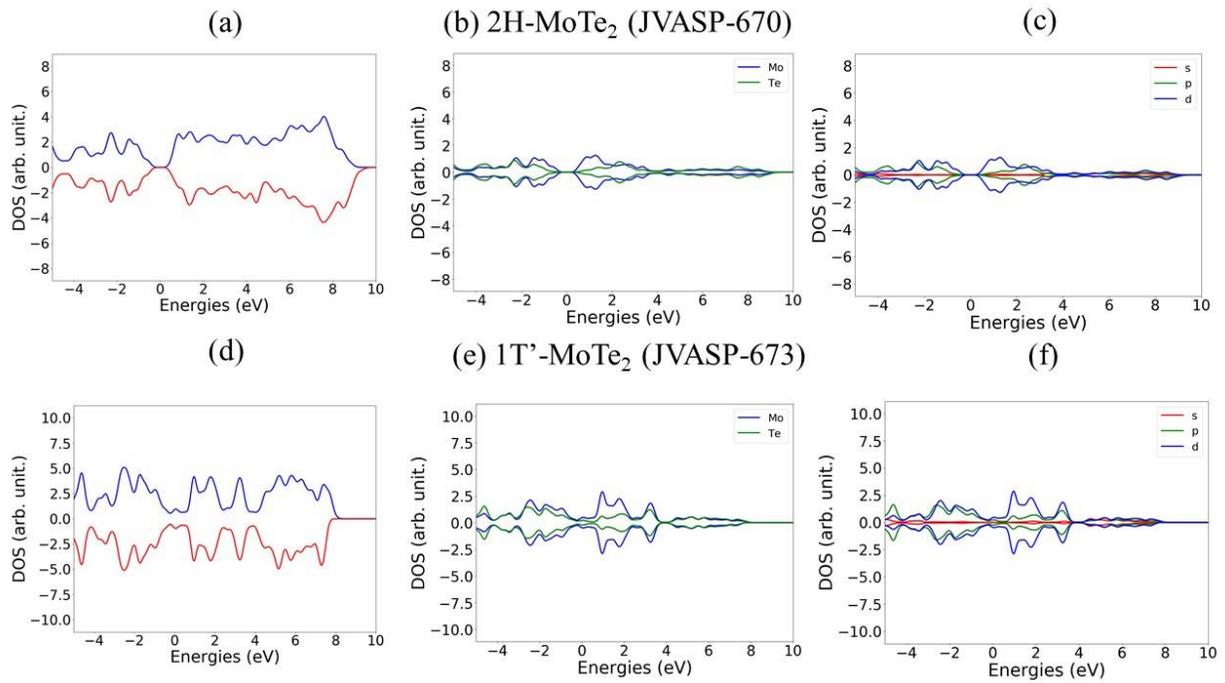

*Fig. S2 Total, element and orbital projected density of states of materials discussed in Fig. 1.*

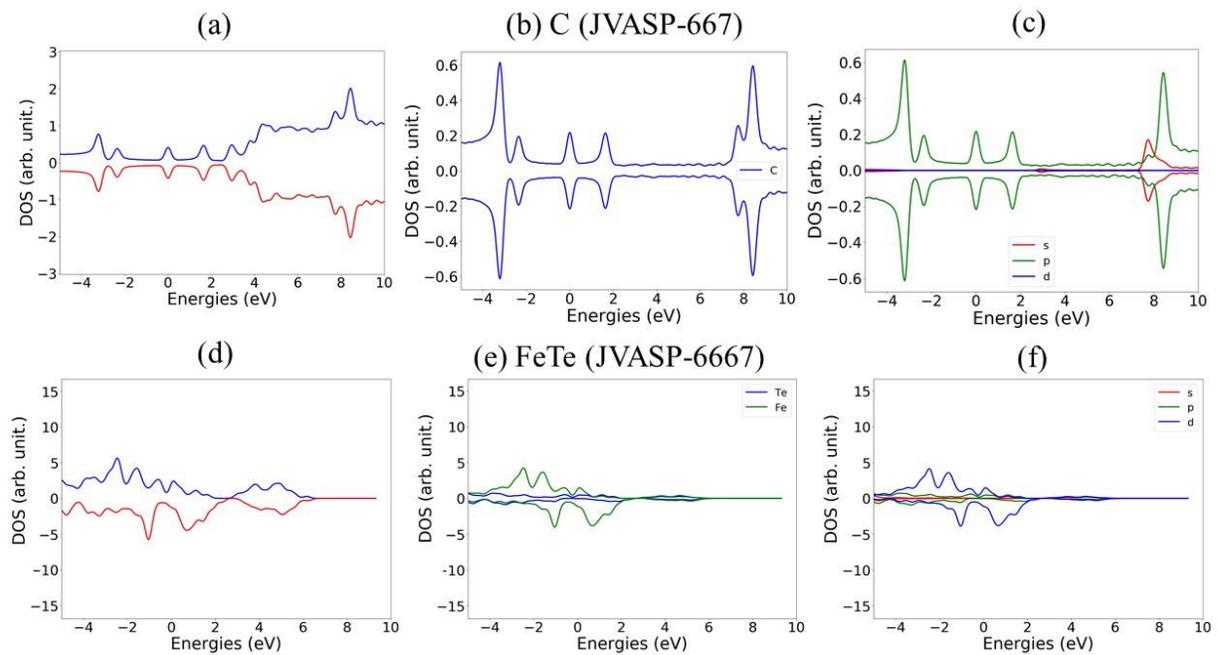

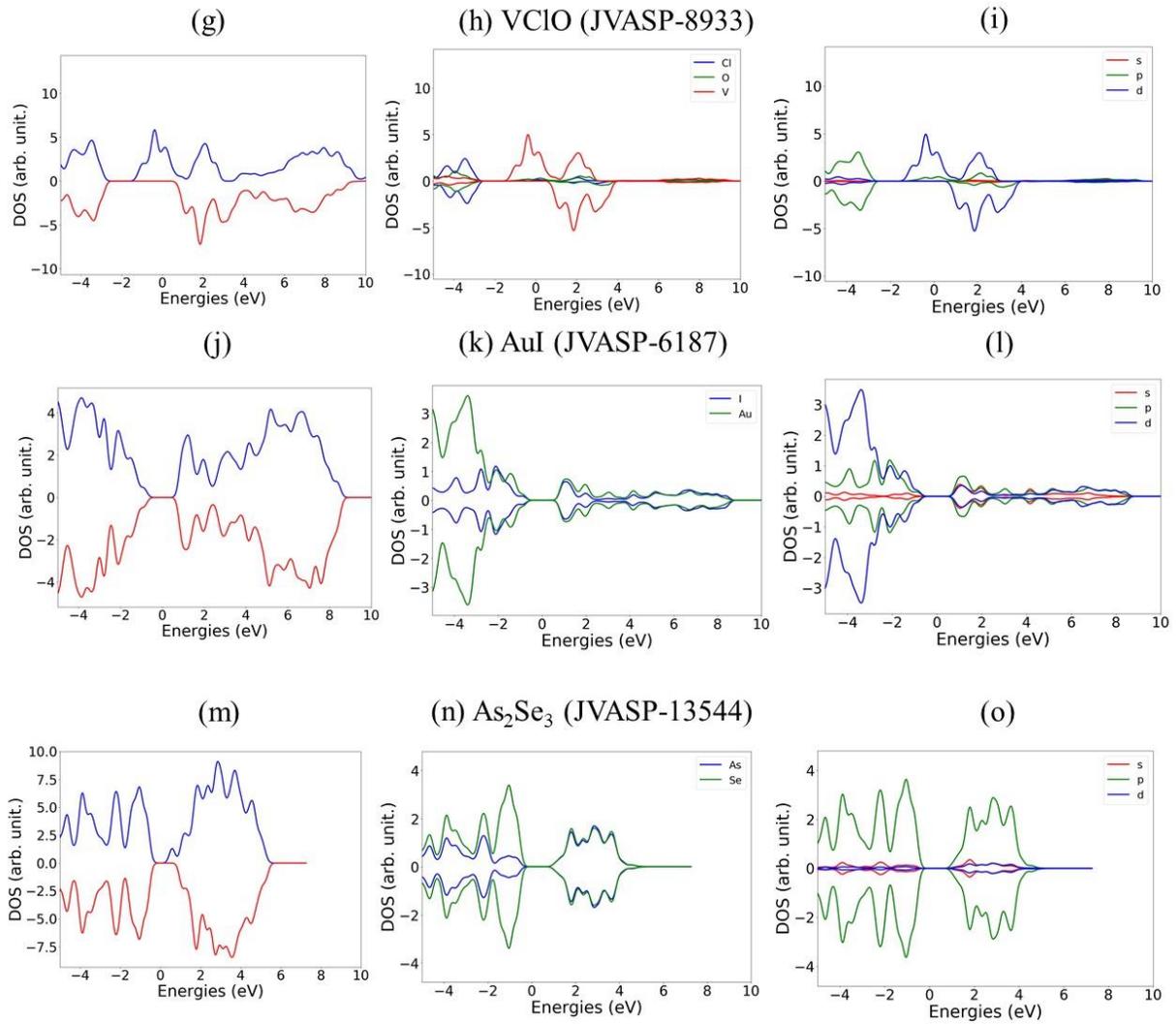

*Fig. S3 Total, element and orbital projected density of states of materials discussed in Fig. 2.*